\documentclass[amssymb,amsmath,pre,twocolumn,superscriptaddress]{revtex4-2}
 
\usepackage{graphicx}

\usepackage{color}

\begin{document}
\title{Concurrency measures in the era of temporal network epidemiology: A review}

\date{\today}
\author{Naoki Masuda}
\affiliation{Department of Mathematics, State University of New York at Buffalo, USA}
\email{naokimas@buffalo.edu}
\affiliation{Computational and Data-Enabled Science and Engineering Program, State University of New York at Buffalo, USA}
\author{Joel C. Miller}
\affiliation{School of Engineering and Mathematical Sciences, La Trobe University, Australia}
\author{Petter Holme}
\affiliation{Tokyo Tech World Research Hub Initiative (WRHI), Institute of Innovative Research, Tokyo Institute of Technology, Yokohama 226-8503, Japan}

\begin{abstract}
Diseases spread over temporal networks of interaction events between individuals. Structures of these temporal networks hold the keys to understanding epidemic propagation. One early concept of the literature to aid in discussing these structures is concurrency---quantifying individuals' tendency to form time-overlapping ``partnerships''. Although conflicting evaluations and an overabundance of operational definitions have marred the history of concurrency, it remains important, especially in the area of sexually transmitted infections. Today, much of theoretical epidemiology uses more direct models of contact patterns, and there is an emerging body of literature trying to connect methods to the concurrency literature. In this review, we will cover the development of the concept of concurrency and these new approaches.
\end{abstract}

\maketitle

\section{Introduction}

\subsection{Temporal network epidemiology, partnership, and concurrency}

The networks over which sexually transmitted infections spread can, to a high accuracy, be described as a sequence of sexual events (sometimes known as ``interactions'', ``contacts'', or ``encounters'') between pairs of individuals. These interactions are the atoms of time-resolved sexual networks, in that it is not meaningful to divide them into sub-events from an epidemiological point of view. Furthermore, the duration of these events have little impact on disease spreading, so a concise mathematical representation of the sexual contact patterns of a population is as a sequence of events $(i,j,t)$, where $i$ and $j$ are the individuals involved and $t$ is the time of the event. We will call such a sequence of events a \textit{temporal network}~\cite{HolmeSaramaki2012PhysRep,HolmeSaramaki2013book,Holme2015EurPhysJB,HolmeSaramaki2019book,Masuda2020book}. Temporal network epidemiology~\cite{Masuda2013F1000,Masuda2017book} connects the traditional compartmental models of theoretical epidemiology~\cite{Anderson1991book,Andersson2000book,hethcote2000mathematics} to temporal networks that we defined above or related classes of time-varying networks, which both represent time-varying contact patterns between individuals. 

We will primarily focus on sexually transmitted infections for the remainder of this review. Like static network epidemiology~\cite{Keeling2005JRSocInterface,Barrat2008book,Keeling2008book,KissMillerSimon2017book,Pastorsatorras2015RevModPhys}, temporal network modeling becomes particularly useful for sexually transmitted infections. The reason is that for this case, interaction events are relatively well defined~\cite{liljeros2003sexual}. This case contrasts to respiratory infections, where the networks are harder to observe and cross-sectionally denser such that mass-action (i.e., well-mixed population) models incorporating some heterogeneity across individuals may be a better approach.
However, our discussion can be generalized to other infectious diseases and their associated temporal networks if interaction events between individuals are reasonably well-defined.

Many factors determine the likelihood of whether or not one event spreads the disease from one individual to another---the nature of the interaction, the health status of those involved, etc.---but one can say for sure that sexual transmissions can only occur at the times of the events, and between the people involved. The structure of a temporal network determines many aspects of the spread of infections~\cite{andersson1999epidemic,Volz2009JRSocInterface,fefferman2007disease,HolmeControlling2016,holme2015information}. Periodic patterns (circadian rhythms, seasonal variations, etc.) are examples of purely temporal structures that can influence spreading. Heterogeneities in the number of network neighbors are a more network-related, yet influential structure.

\begin{figure}
\includegraphics[width=\linewidth]{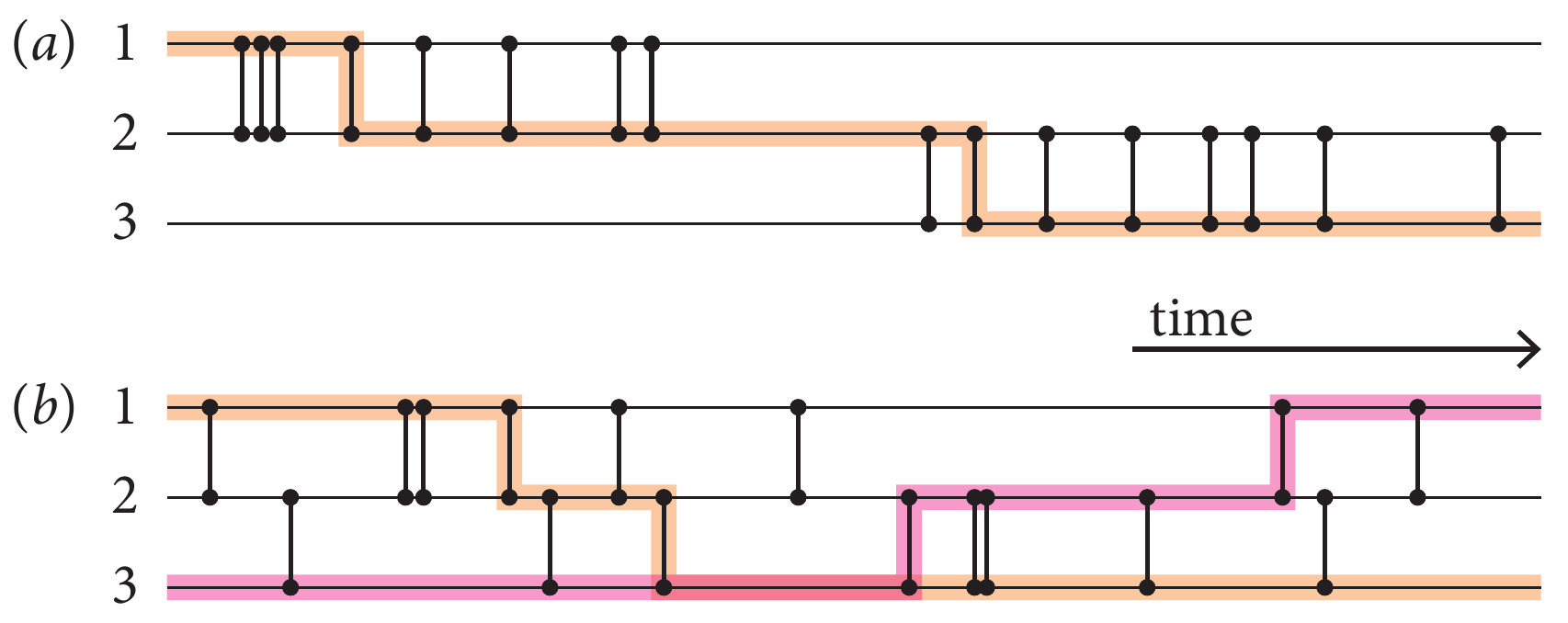}
\caption{Schematic of temporal networks with three nodes that are different in the degree of concurrency. (a) Serial monogamy. (b) Concurrent partnerships.
A vertical bar represents a time-stamped interaction event between a pair of nodes, neglecting the duration of the event. The thicker, shaded lines are examples of time-respecting paths from 1 to 3 and, only in (b), from 3 to 1. This figure is inspired by Fig.~1 of Ref.~\cite{MillerSlim2017PlosOne}.}
\label{fig:Miller}
\end{figure}

Some properties of temporal networks inherently depend on both time and network structure. A fundamental notion of temporal networks is a time-respecting path---whether a path exists from one individual to another through a sequence of events increasing in time. Note that infections can only propagate along time-respecting paths. Suppose that individual 1 is connected to 2 and 2 to 3. If all events between 1 and 2 happen before the events between 2 and 3, there is no time-respecting path from 3 to 1 via 2 (see Fig.~\ref{fig:Miller}(a)). If, on the other hand, the events between 1 and 2 are interspersed with events between 2 and 3, then there are time-respecting paths both from 1 to 3 and from 3 to 1 (Fig.~\ref{fig:Miller}(b)). In the latter case, it is harder to contain an outbreak. The partnerships between 1 and 2 and between 1 and 3 are concurrent in the second scenario (Fig.~\ref{fig:Miller}(b)) but not in the first scenario (Fig.~\ref{fig:Miller}(a)).

Early network studies of disease spreading over sexual events were, in many ways, pioneering. One can argue that Kretschzmar and Morris's Ref.~\cite{Morris1995SocNetw} was the first work of modern computational network science, in the sense that they tuned the network structure and studied the response of a dynamic system on the network. However, these early studies did not consider the temporal network of events as described above. They used a more coarse-grained network of partnerships.

From a modeling point of view, a partnership is a time window, associated with a pair of individuals, within which a disease can spread between them. One typically assumes the likelihood of contagion per unit of time to be constant during a partnership and that partnerships do not take a break and start over again. These early articles rarely state how one could hypothetically reduce the full information of a temporal network of sexual interactions to a network of partnerships---should one casual, non-recurring interaction count as a partnership? Furthermore, papers state a multitude of quantitative definitions of concurrency~\cite{Morris1995SocNetw,Kretzschmar1996MathBiosci,Ghani2000SexTransDis,Mah2010AidsBehav,Morris2010AidsBehav,Unaids2010Lancet,Leung2012TheorPopulBiol,Moody2016AnnEpid,LeeEmmons2019PhysRevE,LeeMoodyMucha2019Springer}, and several studies point out the difficulty of studying  an issue without an agreed definition~\cite{Aral2010CurrInfectDisRep,Lurie2010AidsBehav-limited,Kretzschmar2012AidsBehav,LeeMoodyMucha2019Springer}. Some of the existing confusion and controversy probably stems from the vagueness of the partnership concept. However, concurrency is by now so fundamentally rooted in the theory of sexually transmitted infections that one cannot just ignore it and start anew. Accordingly, a valuable line of research is to connect the temporal network structure to the theory of concurrent partnerships, which is this review's topic.

This review will cover the theory of concurrency in the mathematical literature. In particular, we will try to connect the older literature based on the notion of partnerships with the newer temporal-network-oriented papers. We first describe how concurrency can allow for increased degrees and increased degree heterogeneity, both known to increase disease spread in standard network epidemiology ignoring the dynamic structure. Then, we investigate the impacts of concurrency that cannot be explained without understanding the dynamic structure. We next touch upon the controversy of whether or not high levels of concurrency drive the HIV epidemics in sub-Saharan Africa. However, we do not make a full review of this controversial subject. Finally, we discuss the outlook of temporal network epidemiology and its application to understanding the impact of concurrency.
We stress that our contribution is to formulate concepts and models of concurrency and serve as a tutorial and pointer towards possible future directions, but not as a review of the empirical literature. We only refer to the empirical literature where appropriate to support the conceptual and mathematical frameworks.

\subsection{\label{sub:terminology}Notes on terminology}

In network science, the terminology is often ambiguous. Network epidemiology is no exception. In this section, we state operational definitions for the terms that we use. These are chosen to conform to the literature on both temporal networks and concurrency.

\subsubsection{Event}

An \emph{event} refers to an interaction $(u,v,t)$ between two individuals $u$ and $v$ at time $t$. In general, it is also common to consider the duration of each event in temporal network studies. However, the duration of a single event has no practical meaning for epidemic modeling. For sexually transmitted infections, an event represents a potentially infectious sexual interaction between two individuals.

Typically, one models transmission at an event between an infectious individual $u$ and a susceptible $v$ as a random event with a fixed probability. If one has metadata such as the gender of the individuals or information on condom usage, the probability of infection may depend on them.

The time between two events between $u$ and $v$ is the interevent time~\cite{Karsai2018Springer}. 

\subsubsection{Temporal network}

In this review, we will reserve \emph{temporal network} for a sequence of events $(u,v,t)$ and the set of individuals involved.
Different events may simultaneously occur between different node pairs in a temporal network.

\subsubsection{Partnership}

Given a temporal network, a \emph{partnership} is a set of events between a pair of individuals such that the events are sufficiently evenly distributed. Throughout a partnership, disease could spread from one individual to the other individual at an equal rate. Mathematically we typically represent a partnership by the two individuals involved and its start and end times, which we often assume to be the times of the first and last events between the two individuals, respectively (see Fig.~\ref{fig:kappas} for an example).

\subsubsection{Dynamic partnership network}

Sometimes, we let the edges of a partnership network appear and disappear over time to generate a \emph{dynamic partnership network}~\cite{Dietz1988MathBiosci,andersson1999epidemic}.

\subsubsection{Momentary network}

We refer to the network at a certain time as the \emph{momentary network}. Specifically, 
for dynamic partnership networks, the partnerships existing at time $t$ define the momentary network.
For temporal networks, a node pair $i$ and $j$ forms an edge in the momentary network at time $t$ if and only if there is an event between $i$ and $j$ before $t$ and another event between the same node pair after $t$. These two definitions are identical if we construct a partnership network from the temporal network as we described above, i.e., such that a partnership lasts from the first to the last event between the two individuals. If events are not sufficiently evenly distributed over time during the partnership, the momentary network may not represent the original temporal network data well. 
In other words, the presence of an edge in a momentary network at time $t$ does not mean that there are events along the edge around $t$. Note that this information would typically be lost when representing sexual contacts by a dynamic partnership network.

\subsubsection{Aggregate network}

Finally, we often want to compare epidemic spreading in time-varying networks with that in the counterpart static networks. For a fair comparison, it is necessary to ensure that the different networks under comparison have the same overall number of events. To this end, we use the \emph{aggregate network}, which is defined as the static network in which the weight of  each edge is the same as the fraction of time for which the partnership exists in the given dynamic partnership network. For example, if the observation time window is $t\in [0, 100]$ and $u$ and $v$ are a partnership for $t\in [10, 50]$ only, then the weight of edge $(u, v)$ in the aggregate network, which is static and exists for $t\in [0, 100]$, is equal to $0.4$.

\section{Concurrency as a large mean degree of the network}

In network epidemiology in general, higher degrees (more neighbors) in a contact network signals an easier spread of disease. This is true both for individuals and for entire networks. If an individual has a high degree, it has a higher chance of getting infected and more opportunities to spread the infection than a low-degree individual~\cite{Ghani1998JRStatistSocA}. If the average degree of a network is higher, an epidemic outbreak would happen more easily and be more severe than on a sparser network~\cite{Keeling2005JRSocInterface}. The average degree is also a typical control parameter in studies of component size distributions \cite{Moore2000PhysRevE-exact,Newman2002PhysRevE-transmissibility}; if a component of the network is large, then an infectious disease may spread on a large scale within it.

The reasoning above applies to a scenario where the network is constant throughout the epidemic scenario in consideration. If the network changes over a similar time scale to the epidemics, what static network is most relevant is a challenging question~\cite{Holme2013PlosComputBiol}. Traditionally, the concurrency literature has assumed momentary networks and disease spreading faster than the partnership dynamics~\cite{Morris1995SocNetw,Kretzschmar1995JBiolSyst}. However, this approximation fails to capture a full temporal network picture and thus some outbreak scenarios.

We start by analyzing an early model of concurrent relationships in which increasing concurrency increases the typical degree. We assume undirected networks although contagion is asymmetric for some sexually transmitted infections (e.g., HIV spreads easier from men to women than vice versa~\cite{Patel2014Aids}). This assumption is for simplicity and facilitates model comparison.

In the context of HIV/AIDS, Watts and May carried out a mathematical analysis of a mean-field-type ordinary differential equation (ODE) model of epidemic spreading in their seminal 1992 study ~\cite{Watts1992MathBiosci}. This paper is one of the earliest mathematical papers to discuss the concept of concurrency; see Ref.~\cite{DietzHadeler1988JMathBiol} for an earlier mathematical modeling that focused on monogamy and pair formation and dissolution (i.e., no concurrency) as opposed to well-mixed populations (i.e., concurrency). Their model is a variant of the susceptible-exposed-infectious-recovered/removed (SEIR) model and explicitly incorporates a time delay between a sexual encounter causing transmission and the eventual transition to being infectious as well as the probability that an edge formed continues to exist for a given time. They set the rate at which a susceptible individual is infected, denoted by $\pi(t)$, where $t$ is the time, to
\begin{widetext}
\begin{equation}
\pi(t) = c \beta \left[ \frac{I(t)}{S(t)+E(t)+I(t)} + \pi(t-T) S(t-T) \int_{-\infty}^t \frac{e^{-(t-t')/\tau}}{S(t') + E(t') + I(t')} \text{d}t' \right],
\label{eq:Watts-May}
\end{equation}
\end{widetext}
where $c$ is the rate of acquiring new sexual partners,
$\beta$ is the probability that a susceptible individual is infected by an infectious partner over the duration of the relationship, $S(t)$ is the fraction of susceptible individuals at time $t$, $E(t)$ is the fraction of exposed (infected but not yet infectious) individuals, $I(t)$ is the fraction of infected and infectious, $T$ represents the (fixed) duration of the exposed period, and $\tau$ is the average duration of partnership.

Equation~\eqref{eq:Watts-May} is based on the following explicit and implicit assumptions.  First, those who have developed AIDS, corresponding to the recovered/removed (R) state in the SEIR model, are not sexually active. Note that $S(t)+E(t)+I(t)$ is not generally equal to one. In the first term on the right-hand side of Eq.~\eqref{eq:Watts-May}, the fraction $I(t)/\left[S(t)+E(t)+I(t)\right]$ is equal to the probability that a new partner is infectious. This term represents the rate of acquiring infection from new sexual partners being formed at time $t$. The second term on the right-hand side represents the rate of acquiring infection from an existing partner who acquired infection at time $t-T$ and thus becomes infectious at time $t$.

Second, the first term implicitly assumes that if a newly formed partnership between a susceptible and an infectious individual will eventually transmit, it does so as soon as the partnership forms.  The second term implicitly assumes that if a partner transitions to infectious during a partnership, and a transmission would occur, it does so as soon as the partner transitions to infectious. That is, if infection eventually occurs in the partnership, the transmission is assumed to happen as soon as the partnerhsip begins or the partner becomes infectious.   

Third, the dissolution of partnership obeys a Poisson process. Equivalently, $e^{-(t-t')/\tau}$ in Eq.~\eqref{eq:Watts-May} represents the probability that the partnership that formed at time $t'$ remains at time $t$. 

Fourth, although partnerships are assumed to have exponentially distributed duration, the infection probability is the same for all partnerships, independently of the age of the partnership. 

Suppose that the exposed period $T$ is short compared to the typical partnership duration $\tau$. If a partner $v$ of the focal susceptible individual $u$ is incubating infection (has status E), then because $T$ is small compared to $\tau$ it is likely that $v$ is still in contact with whoever transmitted to $v$. Thus, the second term on the right-hand side as a whole represents the infection of $u$ due to a concurrent partnership of $v$. The authors of Ref.~\cite{Watts1992MathBiosci} essentially varied $\tau$ in their analysis to show that a larger $\tau$, corresponding to stronger concurrency, enhances epidemic spreading. 

However, Eq.~\eqref{eq:Watts-May} indicates that $\pi(t)$ monotonically increases with $\tau$, given that $\beta$ and $c$, which are parameters that control the infection rate per partnership and the rate of pair formation, respectively, are held constant. An increase in $\tau$ implies that a partnership lasts longer, which contributes to the increase in the degree of the individual (i.e., the number of edges, or equivalently, neighbors, that the individual has) averaged over time. The authors also state that $c \tau$ is equal to the number of partners of an average individual~\cite{Watts1992MathBiosci}, i.e., the mean degree over the nodes in the network. Because a larger mean degree is well known to enhance epidemic spreading with other things being equal, their result that epidemic spreading is enhanced by an increase in $\tau$ can be parsimoniously understood as an effect of an increased mean degree~\cite{Bauch2000ProcRSocLondB}.

Many modeling studies that investigate the effect of concurrency, by extending the Watts-May model \cite{Gurski2016MathBiosci} or otherwise \cite{Dietz1992AidsEpid,Chick2000MathBiosci,Eames2004MathBiosci,Leng2018Epid}, fall in the same class; enhanced epidemic spreading in those models can be construed as a consequence of the heightened mean degree of the network.

\section{Concurrency as heterogeneity in the momentary degree distribution}

In static partnership networks, it is known that increasing the heterogeneity in degree can significantly increase the early growth of an epidemic. This is because the first nodes to become infected tend to be those nodes with a higher degree than typical nodes.  These nodes then cause more infections than typical nodes. We now explore some models and concepts of concurrency in which the concurrency allows for increased heterogeneity.

Many early investigations into concurrency held the mean degree fixed and examined the effect of different degree distributions on the epidemic spread. Kretzschmar and Morris pioneered concurrency measures that went beyond the dependence on the mean degree \cite{Morris1995SocNetw,Kretzschmar1996MathBiosci}. Their analysis applies to the momentary network, i.e., those pairs who, at a given time, had events before and would have events again (see Sect.~\ref{sub:terminology} for momentary networks). They studied epidemic process models on top of dynamic network models in which pairs form and dissolve across time and measured the relationship between the epidemic dynamics (such as the fraction of infected nodes and the speed at which infection spreads) and the concurrency measures.

They proposed that, when edges share a node, i.e., if a node has degree $k$ larger than one, then concurrency is present. Otherwise, if $k=0$, with which the node is isolated, or $k=1$, with which the node is in a non-concurrent relationship with one other node, concurrency is absent, at least around the focal node. Let the number of nodes be $N$, and denote the degree distribution of the network by
$\{p(k) : k=0, 1, 2, \ldots \}$; a fraction $p(k)$ of the nodes has degree $k$. The mean degree of the network, denoted by $\langle k\rangle$, is given by $\langle k\rangle = \sum_{k=0}^{\infty} k p(k)$.
Their first measure of concurrency, denoted by $\kappa_1$, is the mean degree, where the mean is taken over the nodes with degree at least one \cite{Morris1995SocNetw} (the same authors defined the reciprocal of this quantity as $\kappa_1$ in their second paper on the topic, Ref.~\cite{Kretzschmar1996MathBiosci}). One obtains
\begin{equation}
\kappa_1 = \frac{\langle k\rangle}{1-p(0)}.
\label{eq:kappa_1}
\end{equation}
Note that $p(0)$ is the fraction of the isolated nodes. A larger $\kappa_1$ means more concurrency. If edges are concentrated in a small fraction of nodes, which leads to a small value of $1-p(0)$, the concurrency is large. The so-called concurrency index, denoted by $\kappa_{I}$, proposed later \cite{Leung2012TheorPopulBiol}, is equal to $\kappa_1 - 1$ if we identify the degree distributions used in the different studies.

For field measurements, UNAIDS Reference Group on Estimates, Modelling and Projections, Working Group on Measuring Concurrent Sexual Partnerships recommended using the fraction of the population that is neither isolated nor in just a single relationship as a concurrency measure \cite{Unaids2010Lancet}. This measure is equal to $\sum_{k\ge 2}p(k) = 1 - p(0) - p(1)$. This measure also enumerates the nodes with concurrent relationships (i.e., $k \ge 2$).

Kretzschmar and Morris also defined a second measure of concurrency, denoted by $\kappa_3$, which they mainly used in their papers (rather than $\kappa_1$) \cite{Kretzschmar1996MathBiosci}. Although they also defined another measure $\kappa_2$ \cite{Morris1995SocNetw,Kretzschmar1996MathBiosci}, it is a rescaled version of $\kappa_3$, so we do not discuss it here.

\begin{figure}
\includegraphics[width=\linewidth]{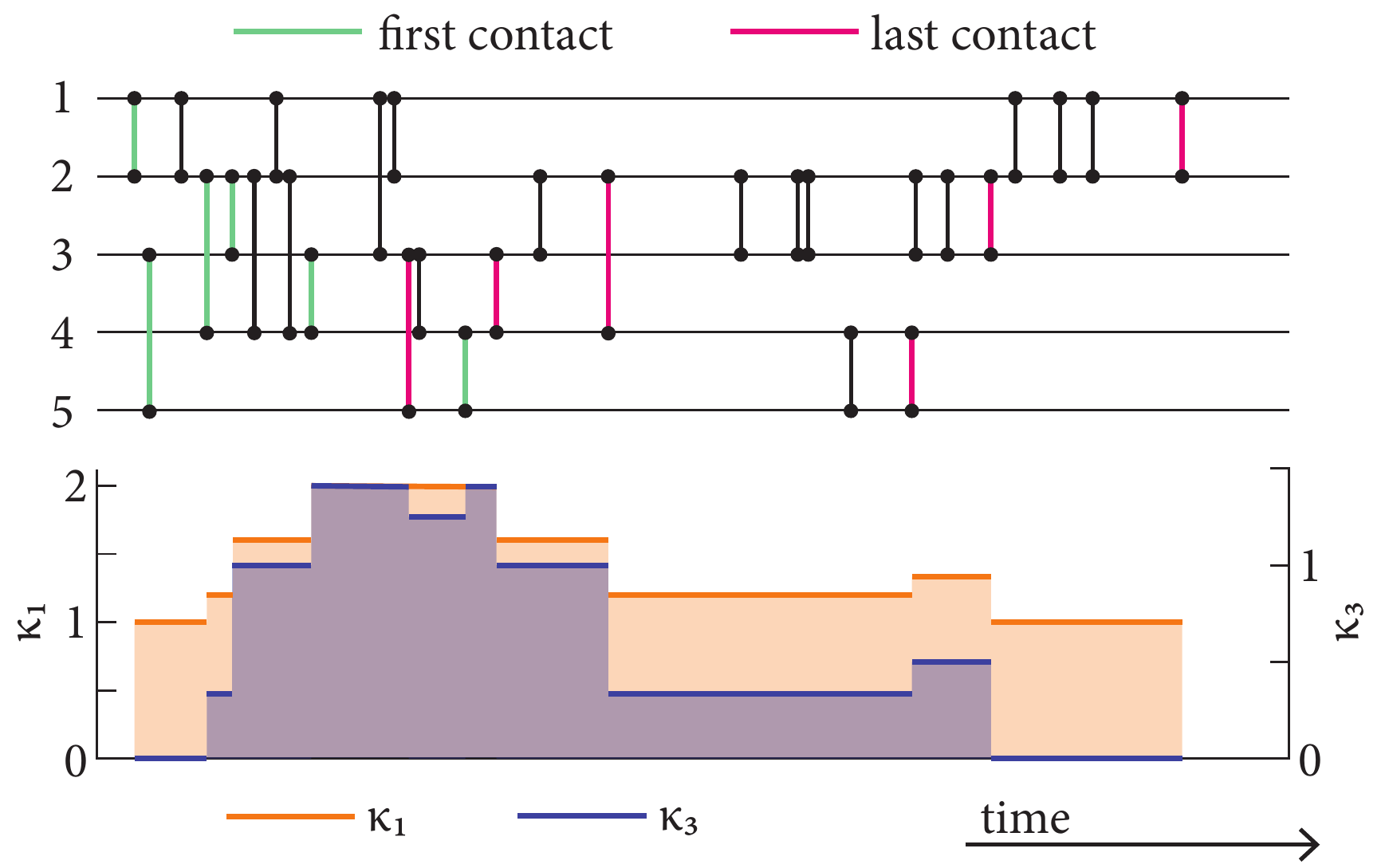}
\caption{An illustration of the time course of the measures $\kappa_1$ and $\kappa_3$ (shown in the bottom panel) in a temporal network (shown by a timeline plot of the events in the top panel). Here we assume that the first event of a pair of nodes is the beginning of a ``partnership'' and that the last event is the end of it, i.e., we consider the momentary network at the given time.}
\label{fig:kappas}
\end{figure}

Suppose that node $v$ has degree $k$, and denote its neighbors by $u_1$, $u_2$, $\ldots$, $u_k$.
Then, each pair of edges, ($v$, $u_i$) and ($v$, $u_j$), where $1\le i\neq j\le k$, are concurrent with each other, increasing the risk of epidemic spreading between $u_i$ and $u_j$ through $v$. There are $k(k-1)/2$ such concurrent edge pairs associated with node $v$. The concurrency measure is defined by the sum of the concurrent edge pairs, over all nodes, i.e., $N\langle k(k-1)/2 \rangle$, divided by the number of edges in the network, i.e., $N\langle k\rangle / 2$, where $\langle \cdot \rangle$ represents the average over nodes. Therefore, one obtains
\begin{equation}
\kappa_3 = \frac{\langle k^2 \rangle}{\langle k\rangle} - 1.
\label{eq:kappa_3}
\end{equation}
A large $\kappa_3$ implies a high level of concurrency. A later defined so-called partnership-based concurrency index, denoted by $\kappa_P$ \cite{Leung2012TheorPopulBiol}, is equal to $\kappa_3 - 1$. An example of $\kappa_1$ and $\kappa_3$ for a temporal network is shown in Fig.~\ref{fig:kappas}.

Another interpretation of $\kappa_3$ goes as follows. If we select a partnership ($u_i$, $v$) uniformly at random, the probability that $v$ has degree $k$ is not equal to $p(k)$ but rather equals $kp(k)/\langle k\rangle$ due to the friendship paradox~\cite{Feld1991AmJSociol}. If $v$ has degree $k$, the number of other nodes $u_j$ ($j\neq i$) that are adjacent to $v$ is $k-1$. Therefore, the expectation of the number of partners that $v$ has other than $u_i$ is
given by
\begin{equation}
\sum_{k=1}^{\infty} \frac{kp(k)}{\langle k\rangle} (k-1) = \kappa_3.
\end{equation}
Under this interpretation, $\kappa_3$ quantifies
the rate at which $u_i$ is at risk of being infected due to $v$ being infected by other neighbors.

\begin{figure}
\includegraphics[width=\linewidth]{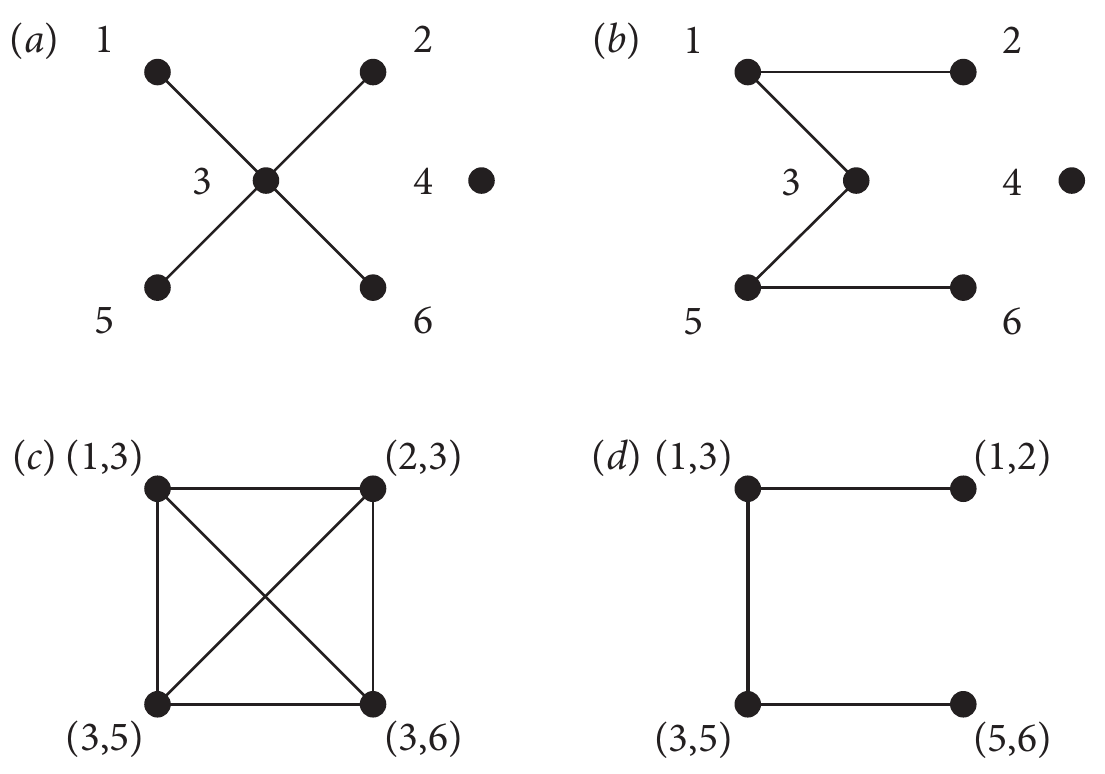}
\caption{A motivating example for the concurrency measure $\kappa_3$ (adapted from Ref.~\cite{Morris1995SocNetw}). Panels (a) and (b) show two networks of currently active partnerships. From a disease spreading point of view, the situation in (a) is worse than the one in (b). This is more evident if one considers the corresponding line graphs in panels (c) and (d). The concurrency measure $\kappa_1$---basically the average degree of the networks---is the same between (a) and (b). However, the concurrency measure $\kappa_3$---basically the average degree of the line graphs shown in (c) and (d), which correspond to (a) and (b), respectively---correctly identifies the first situation as being more densely connected (and thus worse).}
\label{fig:line graph}
\end{figure}

An alternative interpretation of $\kappa_3$ is that it is (up to a constant factor) the average degree of the \emph{line graph} of the original graph~\cite{Enright2018Epid}. By definition, the nodes of the line graph represent the edges of the original graph (i.e., network), and two line-graph nodes are connected if they share a node of the original graph (see Fig.~\ref{fig:line graph} for an example).
 
Numerical simulations of epidemic processes on a temporal network model suggested that the epidemic spread is faster for larger $\kappa_3$, supporting that concurrency contributes to epidemic spreading
\cite{Kretzschmar1996MathBiosci}. This is also the case if $\langle k\rangle$ is kept constant $\kappa_3$ is varied \cite{Morris1997Aids,Morris2000MathPopulStud,Leung2012TheorPopulBiol}.

The concurrency measure $\kappa_3$ and the numerical results obtained in the aforementioned studies are, in fact, consistent with the mean-field theory for epidemic processes on static heterogeneous networks, where heterogeneity refers to that in the node's degree. According to the modified degree-based  mean-field theory, which takes into account heterogeneity in the node's degree in a network and that a node $v$ cannot infect the neighbor that originally infected $v$,
the epidemic threshold for the susceptible-infectious-recovered (SIR) model in terms of the infection rate, where the recovery rate is set to unity without loss of generality, is given in Refs.~\cite{Boguna2003chapter,Dorogovtsev2008RevModPhys,Pastorsatorras2015RevModPhys} as follows:
\begin{equation}
\beta_{\rm c} = \frac{\langle k\rangle}{\langle k^2\rangle - \langle k\rangle} = \frac{1}{\kappa_3}.
\label{eq:beta_c and kappa_3}
\end{equation}
If $\beta > \beta_{\rm c}$, the final epidemic size is large with a positive probability. Equation~\eqref{eq:beta_c and kappa_3} indicates that a large $\kappa_3$ implies that epidemic spread is facilitated in the SIR model on networks. 

The effect of $\kappa_3$ on epidemic spread can be understood in terms of concurrency allowing for more heterogeneous degree distributions which impacts epidemic spread rather than in terms of how partnerships overlap across time. 
Note that $\kappa_3$ was proposed and at least numerically investigated in the mid 1990s \cite{Kretzschmar1996MathBiosci}, preceding seminal studies of epidemic processes in networks with heterogeneous degree distributions (e.g., \cite{Pastorsatorras2001PhysRevLett,Moreno2002EurPhysJB,Boguna2003chapter}). However, the effects of such heterogeneities were known in the field at the time. One example of an early way of dealing with degree heterogeneities is to multiply the basic reproductive number by the following correction factor due to Anderson and May~\cite{Anderson1991book,liljeros2003sexual}:
\begin{equation}\label{eq:correct}
    1+\frac{\langle k^2\rangle-\langle k\rangle^2}{\langle k\rangle} .
\end{equation}

The authors of Refs.~\cite{Morris1995SocNetw,Kretzschmar1996MathBiosci} also acknowledge that $\kappa_3$ is similar to the so-called effective contact number/rate, which uses $\langle k^2\rangle / \langle k\rangle$ to quantify the effect of heterogeneous contact rates in a population on spreading of HIV/AIDS \cite{Anderson1986ImaJMathApplMedBiol,MayAnderson1987Nature,May1988PhilTransRSocLondSerB} (see Ref.~\cite{Nold1980MathBiosci,Hethcote1984LectNotesBiomath} for qualitatively the same results for gonorrhea transmission modeled by the susceptible-infectious-susceptible (SIS) model). Other scenarios such as the fraction of isolated individuals (as investigated through varying $\kappa_1$, thus controlling the frequency of polygamous partnerships) and assortative mixing, with which high-degree nodes tend to connect to each other, were also investigated in the same study \cite{Kretzschmar1996MathBiosci}. These factors can also be mapped to the structure of static networks; for example, the degree assortativity \cite{wylie2001patterns,Newman2002PhysRevLett,Newman2003PhysRevE_mixing}. 

Other studies investigated the effect of concurrency in similar manners. In other words, they simulated or mathematically analyzed epidemic process models in which node pairs form and dissolve over time according to some rules. Then, by keeping the mean degree, $\langle k\rangle$, fixed, either precisely or statistically, they varied model parameters, which changed the degree distribution and hence the value of the concurrency measure used, to see how the extent of epidemic spreading changed \cite{Kretzschmar1994chapter,Kretzschmar1995JBiolSyst,Ghani2000SexTransDis,Doherty2006SexTransDis,Kamp2010PlosComputBiol,KimRiolo2010Epid,Morris2009AmJPublHealth,Eaton2011AidsBehav,Goodreau2012AidsBehav,Leung2015Aids,Leung2017Epid} (also, Fig.~1 in \cite{Garnett1997Aids} is a succinct example to contrast networks with the same $N$ and the mean degree, while the level of concurrency is different). Overall, most of these studies suggest that an increased concurrency, as measured by $\kappa_3$ or otherwise, causes an increase in the size of epidemic spreading. For example, Bauch \& Rand considered an SIS model in which partners form and dissolve dynamically \cite{Bauch2000ProcRSocLondB}. In the model, two isolated individuals form a partnership at rate $\rho/N$, and two individuals form a partnership at rate $\rho \theta / N$ if either of the individuals already has a different partner. Parameter $\theta (0\le \theta\le 1)$ controls the level of concurrency; if $\theta=0$, all the partnerships are monogamous, i.e., no node has a degree above one. If $\theta$ is larger, larger degrees are allowed. Finally, any partnership is assumed to break up at rate $\sigma$. They analytically calculated the degree distribution in the equilibrium, which is a Poisson distribution when $\theta=1$ and has a thinner tail when $\theta<1$. The derived degree distributions led to $\kappa_3 = \rho \theta / \sigma$ \cite{Bauch2000ProcRSocLondB}. They also calculated the mean degree, which we do not show here because the expression is complicated. They then investigated the effect of $\kappa_3$ on the final epidemic size and time evolution of the fraction of infected nodes, by keeping the mean degree constant. They showed that the final epidemic size increases with $\kappa_3$ in most cases. 

\section{Concurrency as a temporal property of networks}

Most of the numerical and analytical results, including those reviewed in the previous sections, employed dynamic network models to investigate the effect of concurrency on epidemic spreading. This is valid because concurrency is a temporal notion. However, the effects of concurrency claimed by many of these papers are already expected from our understanding of static network epidemiology (i.e., how the structure of the static network varies as a concurrency parameter varies, which then affects how infections spread in the static network). Specifically, various previous results stated in terms of $\kappa_1$, $\kappa_3$, or similar measures are effectively restatements of the foundational results of mathematical epidemiology and network science without referring to concurrency, i.e., a large mean degree or a high heterogeneity in the degree distribution given the mean degree yields an increased level of epidemic spreading in networks. This raises the question of whether concurrency on its own can affect how infection spreads or whether it only acts indirectly by affecting node degrees.  More recent analyses that attempt to single out the effect of concurrency from that of static network structure say that it has an important role.

A consensus is that $k\ge 2$ for a node implies the presence of concurrency and $k\le 1$ implies its absence. A majority of concurrency studies using mathematical/computational models explicitly or implicitly assume that the degree distribution is measured at a certain point of time, i.e., for the momentary network (section~\ref{sub:terminology}). Edges in a momentary network at time $t$ thus represent pairs of nodes that had an event in the past and will have one again in the future, so that the partnership is ongoing at time $t$. Therefore, the momentary degree larger than one can be used as evidence of concurrency \cite{Morris2009AmJPublHealth}, and one can quantify this using, for example, $\kappa_1$ or $\kappa_3$. Asking whether the momentary degree is larger than one is also used in survey studies \cite{Morris2010PlosOne}. Then, studies using a dynamical network model would involve a parameter to control the level of concurrency (e.g., the propensity that nodes are polygamous, i.e., $k\ge 2$ at any given point of time) and examine how epidemic spreading changes as one varies the control parameter. The degree distribution in the equilibrium or averaged over time has mostly been used for quantifying the level of concurrency. However, there are different sequences of momentary networks that, when aggregated across time, produce the same static network. These different sequences can vary dramatically in how concurrency appears, and the static network structure cannot explain the resulting changes in how epidemics spread. These are fundamentally due to concurrency and not merely because concurrency facilitates other effects.

To make this point clearer, in the following sections, we survey some recent results that explicitly aimed to single out the effect of concurrency without being confounded by differences in aggregate networks.

\subsection{Overlap of time windows of edge activation} 

Lagarde \emph{et al.}\ defined a concurrency measure for each node, called the individual index of concurrency (\emph{IIC}), as follows \cite{Lagarde2001Aids}. Consider two edges incident to node $v$, denoted by $e_1$ and $e_2$. By considering HIV/AIDS, which their study is based on, we assume that a sexual partnership is formed on $e_1$ and $e_2$ during a time window $[t_1^{\rm start}, t_1^{\rm end}]$ and  $[t_2^{\rm start}, t_2^{\rm end}]$, respectively. Note that the following definition can be easily generalized to the case in which $e_1$ or $e_2$ is activated in multiple time windows. Then, we denote the overlap of $[t_1^{\rm start}, t_1^{\rm end}]$ and  $[t_2^{\rm start}, t_2^{\rm end}]$ by $d$. Specifically, $d$ is the length of time for which both edges are activated, thus concurrent. Because the two time windows of edge activation may overlap even if they occur at random times, they calculated the expected size of the overlap when the two time windows are independently and uniformly randomly located in terms of the time, which we denote by $\epsilon$. If $d/\epsilon = 1$, there is no excess concurrency between $e_1$ and $e_2$ relative to the uniformly random case. If $d/\epsilon > 1$, there is concurrency beyond randomness. Because $0 \le d/\epsilon < \infty$, they defined $r = (d/\epsilon - 1)/(d/\epsilon + 1)$ such that $-1\le r < 1$; the uniformly random overlap corresponds to $r=0$; higher overlap yields $r>0$; and lower overlap yields $r<0$. They summed $r$ over all edge pairs incident to node $v$ to define $v$'s concurrency, i.e., IIC. They performed an interview study in five cities in Africa with an observation time window of one year. They found no correlation between the IIC value and whether or not the sampled people are HIV-infected.  
 
Assuming that the event time and its duration are randomly generated and independent for different edges, different authors defined the temporal coherency as the probability that the time windows of activation on two edges overlap \cite{Moody2016AnnEpid,LeeEmmons2019PhysRevE}. For empirical data of temporal networks, the same quantity can be measured as the fraction of edge pairs that have overlapping active time, where the edge may be optionally assumed to be active all the time between its first and last event times \cite{LeeMoodyMucha2019Springer}. The temporal coherency is similar to a normalized variant of $d$. Note that one can consider both IIC and temporal coherency for individual nodes or the entire network (as the average of the quantity over all the nodes or all the edge pairs in the network).

An important quantity for interpreting how concurrency impacts disease spread is the reachability of a network. The reachability is defined for a temporal network as the expected fraction of node pairs $i$ and $j$ with a time-respecting path from $i$ to $j$ starting at a random time between the beginning and end of the data \cite{Moody2002SocForces,Holme2005PhysRevE,Lentz2013PhysRevLett} (also see \cite{Armbruster2017NetwSci} for a mathematical analysis). The reachability of the temporal network generally increases as the temporal coherency increases \cite{Morris2009AmJPublHealth,Moody2016AnnEpid,LeeEmmons2019PhysRevE,LeeMoodyMucha2019Springer}. Concurrency implies that two edges sharing a node simultaneously exist. For example, if edge $(1, 2)$ and $(2, 3)$ are simultaneously active, then during this period $3$ is reachable from $1$ and vice versa. If the two edges are activated in non-overlapping times, then either $3$ is not reachable from $1$ or vice versa. This is why reachability is expected to increase as the temporal concurrency increases. By definition, the reachability ignores infectivity and therefore does not directly translate into observed epidemics. However, since high average reachability is positively correlated with the final outbreak size, the results in Refs.~\cite{Morris2009AmJPublHealth,Moody2016AnnEpid,LeeEmmons2019PhysRevE,LeeMoodyMucha2019Springer} are consistent with the common claim that the concurrency positively contributes to the severity of outbreaks.

\subsection{Epidemic model with a careful control of concurrency}

Related to the reachability argument, let us again consider two edges that share a node, $(1,2)$ and  $(2,3)$. In Ref.~\cite{MillerSlim2017PlosOne}, the authors pointed out that indirect transmission of infection between $1$ to $3$ through $2$ is possible if events on the two edges occur concurrently. Figure~\ref{fig:Miller}(b) shows such a case. In this figure, although the exact timings of the events generally differ on the two edges, the events on them are concurrently occurring on a coarser time scale of the entire observation time window shown in the figure. In contrast, if events on the two edges are not concurrent, the indirect disease transmission may happen in one direction (e.g., from $1$ to $3$) but not in the opposite direction, at least within a time window of interest (see Fig.~\ref{fig:Miller}(a)). Furthermore, disease transmission can go faster in the concurrent case than the non-concurrent case. With reference to Fig.~\ref{fig:Miller}, this is because the pathogen that has transmitted from $1$ to $2$ can almost immediately travel to $3$ in the concurrent case, but it must wait in the non-concurrent case \cite{MillerSlim2017PlosOne}. Crucially, this discussion compares scenarios that are different in terms of concurrency (no matter how one measures it) but share the static network structure including the weight (i.e., number of events) of each edge; compare the two temporal networks shown in Fig.~\ref{fig:Miller}. The networks shown in Fig.~\ref{fig:Miller} are small and for expository purposes. However, the claim that concurrency mitigates limitations on transmission pathways and therefore enhances the epidemic size and speed has a general value because the same argument holds for larger networks.

In the same study~\cite{MillerSlim2017PlosOne}, the authors examined a susceptible-infectious (SI) model with births and deaths in discrete time. They preserved the structure of the aggregate static network and the weight of each edge in the aggregate network (and hence the degree of each node) and varied the amount of concurrency. Specifically, they assumed a population in which each node had its own degree, and whenever a partnership ended a replacement partner was immediately found so that its degree remained fixed.

To investigate concurrency, the authors of~\cite{MillerSlim2017PlosOne} looked at regular random graphs, where each node has the same degree, $k^{\rm tp}$, at each point of time. In the serial monogamy case, $k^{\rm tp}=1$. To make a fair comparison with polygamy cases, they imposed that each node has degree $k$ in the static network obtained by the aggregation over a time window. This implies that if $k^{\rm tp}$ is small, the nodes have to switch the partners rapidly to collect $k$ partners over time. Note that the concurrency is entirely absent in this case. If $k^{\rm tp}$ is large, then the concurrency is present, and the nodes do not rapidly switch the partners. They also carefully controlled the infection rate parameter so that the comparison across different $k^{\rm tp}$ values is fair, i.e., the weighted aggregate network does not depend on the $k^{\rm tp}$ value. 

They carried out numerical simulations with $k^{\rm tp}$ varied. They found that concurrency typically enhanced the early growth of an epidemic, but it typically had a small impact on the ultimate equilibrium number infected.  

\subsection{Network fluctuations}

\begin{figure}
\includegraphics[width=0.9\linewidth]{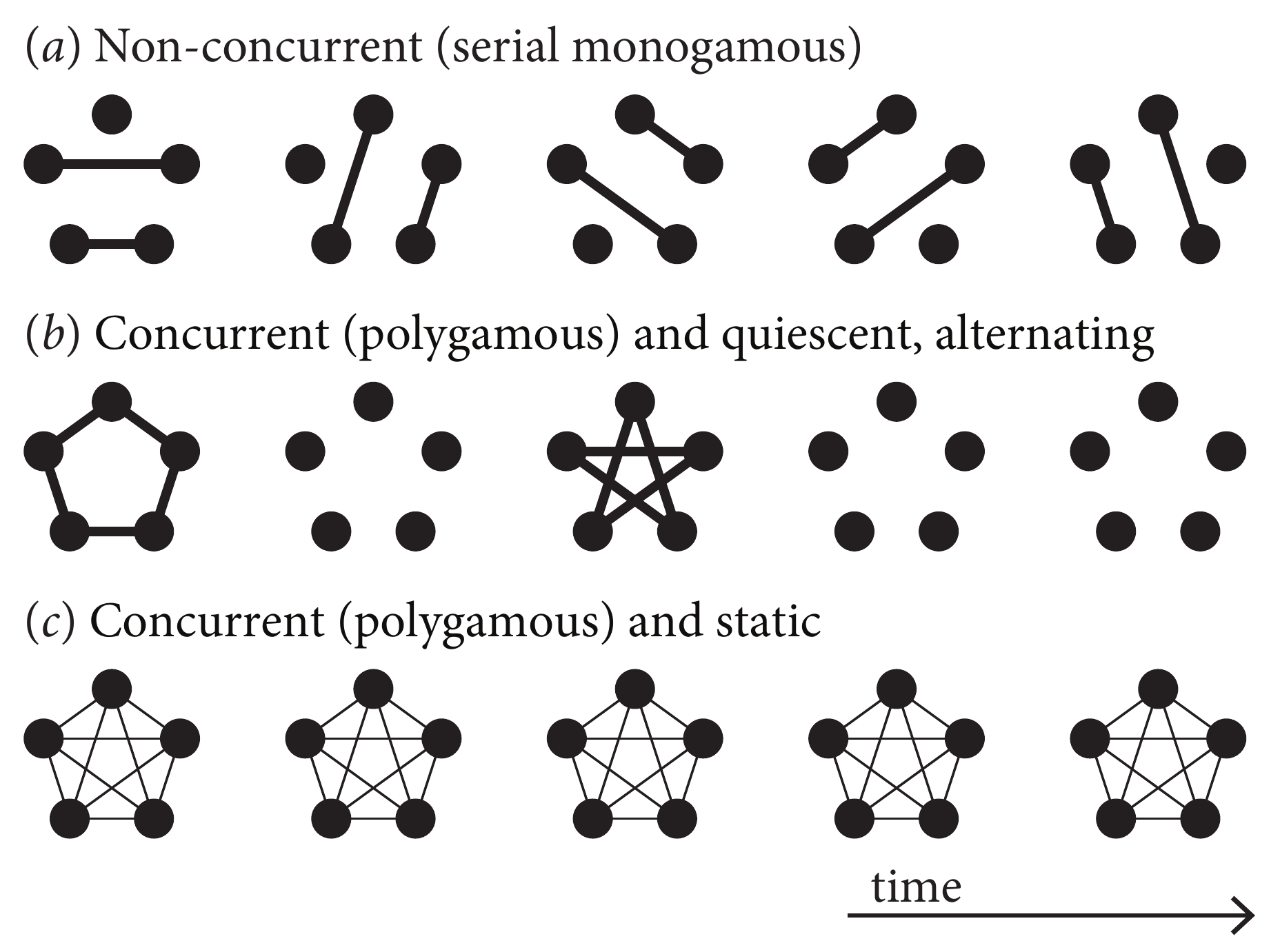}
\caption{Three partnership networks that have the same aggregate network. (a) Dynamic partnership network without concurrency. Each node is either isolated ($k=0$) or involved in a non-concurrent relationship ($k=1$) at each time. (b) Dynamic partnership network with concurrency. At two out of five times, a cycle graph is formed such that all nodes have degree 2. To compensate for a relatively large mean degree at these times, the network at the other three discrete times are the empty network, where every node is isolated. (c) Static partnership network with concurrency. The complete graph is presented. The weight of each edge is $1/5$ of that of the edge shown in (a) and (b). The time-averaged network of the dynamic partnership networks shown in (a) and (b) is the complete network shown in (c).}
\label{fig:concurrency comparison}
\end{figure}

In empirical data of temporal networks, events between nodes are often bursty~\cite{Barabasi2005Nature,Karsai2018Springer}, such that there tends to be a burst of events in some periods and quiescence in others, in a manner not captured by Poisson processes or ODE models~\cite{HolmeSaramaki2012PhysRep,Masuda2013F1000,Holme2015EurPhysJB}. Therefore, individual momentary networks may carry large fluctuations so that their time average or ensemble average does not represent the original temporal network of dynamic partnership network well. Figure~\ref{fig:concurrency comparison} compares three dynamic partnership networks in discrete time that share the same time-averaged aggregate network. Note that we allow partnerships to form and dissolve several times for our discussion. Then, the dynamic partnership network shown in Fig.~\ref{fig:concurrency comparison}(a) completely lacks concurrency because each node is either isolated ($k=0$) or in a non-concurrent relationship ($k=1$) at any discrete time. By contrast, the dynamic partnership network shown in Fig.~\ref{fig:concurrency comparison}(b) has some amount of concurrency. At two out of the five times, each node is involved in concurrent partnerships. At the other three discrete times, the network in Fig.~\ref{fig:concurrency comparison}(b) is empty. This is to make the dynamic partnership networks shown in Figs.~\ref{fig:concurrency comparison}(a) and \ref{fig:concurrency comparison}(b) to have the same amount of partnership averaged over time for each edge. In fact, the partnership is present between each node pair one out of the five times in both networks. The aggregate network for both dynamic partnership networks is the static complete graph shown in Fig.~\ref{fig:concurrency comparison}(c). If we calculate a concurrency measure such as $\kappa_1$ or $\kappa_3$ for the time average of the two dynamic partnership networks, the value will be the same. We set the weight of each edge to be $1/5$ in Fig.~\ref{fig:concurrency comparison}(c). Then, all the three networks shown in Fig.~\ref{fig:concurrency comparison} have the same time-averaged (or equivalently, aggregate) network, which is the complete graph with edge weight $1/5$ (i.e., any of the five time-independent networks shown in Fig.~\ref{fig:concurrency comparison}(c)).

For the network at each discrete time shown in Fig.~\ref{fig:concurrency comparison}(a), we obtain $\kappa_3 = 0$ because $\langle k\rangle = \langle k^2\rangle = 4/5$. Therefore, the time average of $\kappa_3 = 0$. For the non-empty networks in Fig.~\ref{fig:concurrency comparison}(b), we obtain $\kappa_3 = 1$, which follows from $\langle k\rangle = 2$ and $\langle k^2\rangle = 4$. For the empty network in Fig.~\ref{fig:concurrency comparison}(b), $\kappa_3$ is ill-defined since $\langle k\rangle = 0$. However, it makes no sense to assign any value other than $\kappa_3=0$ in this situation. Then the time average of $\kappa_3$ for the dynamic partnership network shown in Fig.~\ref{fig:concurrency comparison}(b) is $2/5$, which is larger than that for the dynamic partnership network shown in Fig.~\ref{fig:concurrency comparison}(a). This result is consistent with our claim above that the dynamic partnership network shown in Fig.~\ref{fig:concurrency comparison}(b) is more concurrent than that shown in Fig.~\ref{fig:concurrency comparison}(a). We emphasize that the calculation of $\kappa_3$ in this case is possible only when we examine dynamic partnership networks individually at different discrete times, which is different from calculating $\kappa_3$ of the aggregate static partnership network. 

Furthermore, because $\kappa_3$ is defined for unweighted networks, strictly speaking, one cannot calculate it for the time-averaged network shown in Fig.~\ref{fig:concurrency comparison}(c). However, it is hard to imagine any other generalization of Eq.~\eqref{eq:kappa_3} to weighted networks than replacing the degree by the node's strength (i.e., the sum of the weights of a node's incident edges). Nevertheless, this straightforward generalization produces a negative value of the concurrency for the static weighted network shown in Fig.~\ref{fig:concurrency comparison}(c), i.e., $\kappa_3 = -4/5$, because $\langle k\rangle = 1/5$ and $\langle k^2\rangle = 1/25$.

In Refs.~\cite{Onaga2017PhysRevLett,Onaga2019Springer}, the SIS model in continuous time is analyzed on dynamic partnership networks that switch from one to another at regular intervals, which we refer to as switching networks. Let us now consider Fig.~\ref{fig:concurrency comparison} as representing network dynamics in continuous time. In switching networks with concurrency, schematically shown in Fig.~\ref{fig:concurrency comparison}(b), the epidemic threshold is smaller (therefore, infection is more likely at least near the epidemic threshold) than in switching networks without concurrency, schematically shown in Fig.~\ref{fig:concurrency comparison}(a), even though the aggregate network is the same in the two cases. Although it is only illustrative, Figs.~\ref{fig:concurrency comparison}(a) and \ref{fig:concurrency comparison}(b) represent the situations in which fluctuations around the mean of the networks at different times are large, such that it is not helpful to approximate the degree distribution of the network at each time by the time average. For the same reason, it may be invalid to average a static network of the daytime with one of the nighttime and analyze the concurrency of the time-averaged network.

\subsection{Time average of concurrency vs concurrency of the time-averaged network}

The amount of concurrency may vary over time. For example, if one looks at the overlap of the time windows of edge activation, the two edges are concurrent when both edges are activated. They are not concurrent if either edge is not active. The latter sounds trivial, but comparison with randomized cases often requires that the original data or model have some periods for which the edges are not active. As another example, in Fig.~\ref{fig:concurrency comparison}(b), the concurrency is high when the time-independent network contains a cycle. At other times, the network is empty, and concurrency is absent.

Some concurrency measures including $\kappa_1$ and $\kappa_3$ are functions of the degree distribution of the network, $\{p(k): k=0, 1, \ldots \}$. A convenient method to calculate such a concurrency measure for a dynamic partnership network model may be to calculate it for the degree distribution in the equilibrium. However, crucially, what we obtain in the equilibrium is not a single degree distribution, but a distribution of the degree distribution. For example, for the network shown in Fig.~\ref{fig:concurrency comparison}(b), the equilibrium is characterized by a two-peak distribution of the degree distribution, i.e., $p(0) = 1$ and $p(k) = 0$ for $k\ge 1$ with probability $3/5$, and $p(2) = 1$ and $p(k)=0$ for $k\neq 2$ with probability $2/5$. The two degree distributions yield different levels of concurrency, e.g., the $\kappa_3$ value, and time aggregation of the concurrency measure, such as a simple time average, tells us how concurrent the entire dynamic partnership network is. It is incorrect to consider the time (or ensemble) average of the degree distribution first (which yields $p(0) = 3/5$, \ $p(2) = 2/5$, and $p(k)=0$ for  $k \notin \{ 0, 2\}$) and then calculate the concurrency measure, or consider the time average of the network first (which yields the weighted network shown in Fig.~\ref{fig:concurrency comparison}(c)) and then calculate the concurrency measure. The same caveat applies to IIC. One calculates the overlap of the time windows of edge activation observed at each point of time, which one sums over the entire observation time window to obtain IIC after further manipulations. We emphasize that the time aggregation of a concurrency measure and the concurrency measure for the time or ensemble average of an evolving network are generally different from each other.
For assessing the effect of concurrency on epidemic spreading, the former is relevant but not the latter, because the time or ensemble average of networks is not the object on which epidemic processes occur.

\section{Other temporal networks analyses related to concurrency}

In this section, we will discuss some further methods to analyze temporal networks inspired by concurrency.

\subsection{Start and end times of edges}

A central assumption in the traditional concurrency literature is that a partnership is a meaningful low-level representation of interactions transmitting disease. This assumption has some support in the temporal network literature. Reference~\cite{Holme2014SciRep} concludes that it is a more relevant simplification of a temporal network to reduce it to a weighted partnership network than to a static network with events generated with the same interevent time statistics as in the original data.   In the wake of Ref.~\cite{Barabasi2005Nature}, a line of research seemed to take for granted that time-stamped event data is well-described as a static network with interevent times sampled from a fat-tailed distribution~\cite{Min2011PhysRevE,VazquezA2007PhysRevLett}. Reference~\cite{Holme2014SciRep} contrasts this ``ongoing link picture'' with a ``link turnover picture'' that rather resembles the pair-formation models of the concurrency literature~\cite{blanchard1990modelling,altmann1995susceptible,Morris1995SocNetw,KRETZSCHMARpairformation}. A temporal network simplified by the link turnover picture contains information about the first and last events between two nodes and the total number of events, but nothing about the interevent times. For disease spreading on empirical temporal networks, it turns out that numerical results for the link turnover picture closely match simulations on the empirical data. The ongoing link picture, on the other hand, tends to overestimate the final outbreak sizes.

\subsection{Reducing temporal networks to networks of concurrent partnerships}\label{sec:reducing}

\begin{figure}
\includegraphics[width=\linewidth]{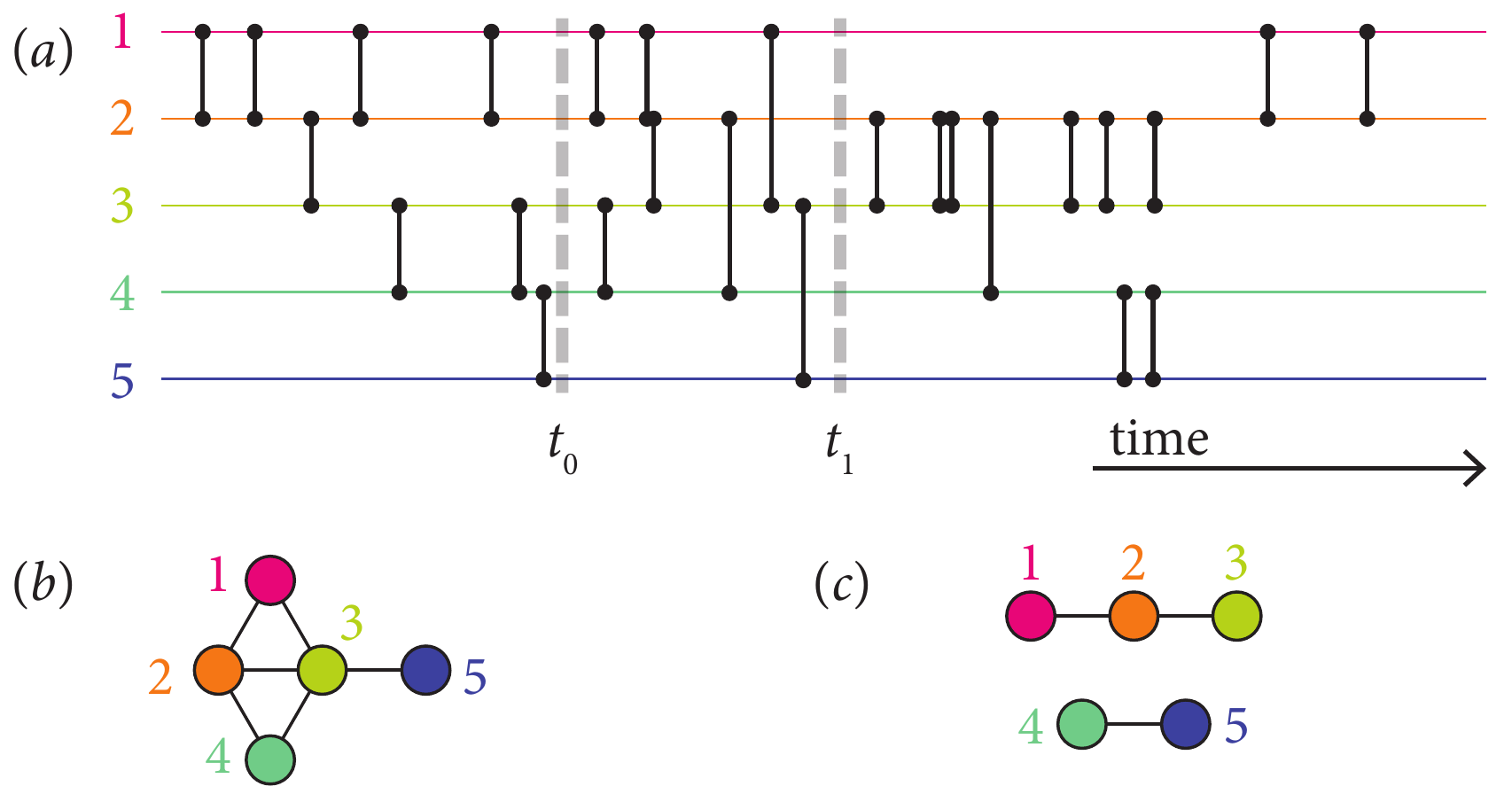}
\caption{Two ways of projecting time out of a temporal network. Panel (a) shows a temporal network. Panel (b) shows the network of nodes that have an event within the time window $[t_0,t_1]$. Panel (c) shows a network of concurrent partnerships---pairs of nodes that have an event before and after $[t_0,t_1]$.}
\label{fig:window_concurrent}
\end{figure}

The typical way to reduce a temporal network to a static network is to connect any node pair by an edge when the two nodes have one or more events within a particular time window~\cite{krings2012effects,sekara2016fundamental,Holme2015EurPhysJB}. In other words, two nodes $u$ and $v$ form an edge $(u,v)$, active from $t_0$ to $t_1$, if there is an event within the interval $[t_0,t_1]$. An alternative way to project out the time from the temporal network is to place edges between pairs of nodes if they have events both before and after the interval. This construction gives concurrent partnerships over the entire interval $[t_0,t_1]$. If $t_0=t_1$, we obtain the momentary network.  Reference~\cite{holme2003network} observed that the degree distributions of momentary networks are closer to power-law distributions than those of aggregate networks. More pertinent to this review, however, is that the network of concurrent partnerships is a worse way to project a temporal network to a static network in the sense that it preserves the ranking of important nodes worse than a standard time-windowed network~\cite{Holme2013PlosComputBiol} (see Fig.~\ref{fig:window_concurrent}).

\section{The debate about the concurrency hypothesis}

The role of concurrency in the HIV epidemics of sub-Saharan Africa has been a highly contentious subject. Some regions of this area have had a hundred-fold higher HIV prevalence than the world average~\cite{Kharsany2016Open}. When the severity of the epidemics in sub-Saharan Africa became apparent in the 1990s, researchers immediately sought explanations within this area's sexual patterns and practices. Concurrency was one such explanation; promiscuity and polygamy were others, along with the belief that sex with virgins could cure HIV~\cite{Inungu}. Around the turn of the millennium, a host of studies of sexual behaviors were published that seemed to suggest that, effectively, general sexual risk behavior was not higher in sub-Saharan Africa than in the rest of the world---see, e.g., Ref.~\cite{Wellings} and further references therein. These new empirical data did not change all theorists' minds, which set off an acrimonious debate that we will attempt to give a flavor of in this section.

One side of this polemics asserts that: First, a higher concurrency characterizes sub-Saharan Africa's sexual act patterns than in the West and elsewhere in the world. Second, simulation studies seem to show that concurrency is an essential factor for HIV dynamics. Because of these two reasons, concurrency is the key driver of the HIV epidemics in these countries~\cite{Mah2010AidsBehav}. The other side claims that concurrency does not matter much, and more likely explanations of the higher prevalence are co-infections increasing the susceptibility or non-sexual transmission pathways~\cite{Sawers2010JIntAidsSoc,rothenberg}.

One seminal paper arguing against the concurrency hypothesis was Ref.~\cite{Lagarde2001Aids} by Lagarde and coauthors. This was the first survey to record both concurrencies---using their metric, IIC, that we discussed above---and HIV status. The authors interviewed 9,643 persons in five regions of sub-Saharan Africa. They found no correlation between their concurrency metric and whether or not a person was HIV positive.

This study was subsequently criticized, mostly due to alleged methodological flaws. Reference~\cite{rothenberg} notes that Ref.~\cite{Lagarde2001Aids} measures current concurrency only, but HIV positive must have been infected earlier. Reference~\cite{Morris2010PlosOne} argues that ``the predicted empirical signature of concurrency's effect on the transmission is not a correlation between index case concurrency and their own HIV status, but a correlation between index case concurrency and their partner's HIV status''. 

To illustrate this argument, consider the scenario of an individual $u$ having partners $v$ and $w$ in a time period in which $v$ and $w$ have no other partners. The risk to $u$ depends on the number of events with $v$ and $w$, but the ordering of those events has no effect. However, the probability that $u$ serves as a conduit of infection between $v$ and $w$ increases if these events are interspersed. This effect is magnified if we account for the fact that $u$ would have the highest viral load in the acute phase shortly after infection~\cite{hollingsworth2008hiv}. So even though there is a correlation between the infection statuses of partners in almost any compartmental model on networks~\cite{cai_solving}, one person's concurrency and his/her HIV status could still be uncorrelated according to Ref.~\cite{Morris2010PlosOne}.

A decade later, another major survey---this time longitudinal, geolocated, and accounting for the possibility that current concurrency may affect future HIV status---did not find an association between concurrency levels in communities and HIV acquisition among women~\cite{tanser}. The pattern repeated with papers from both sides accusing the other of withholding information, misinterpreting results, and committing methodological errors~\cite{sawers_isaac,Lurie2010AidsBehav-limited,Epstein2011JIntAidsSoc,Morris2011Lancet}. 

A potential explanation of some of these observations is found in Ref.~\cite{MillerSlim2017PlosOne}, which showed that in many cases concurrency can play a large part in the early growth of an epidemic even while it plays little role in the eventual equilibrium reached. So the apparent observation from some models that the growth of the epidemic can only be reproduced by incorporating concurrency may still be consistent with measurements taken in long-established epidemics which measure a similar incidence of infection in populations with different levels of concurrency. There are plenty of further twists and turns in this debate about the ``concurrency hypothesis'' that we will not dwell on further. We will not call a winner of this debate.

\section{Outlook}

Temporal networks do not capture the full truth about how contact patterns affect epidemics. They do not typically, for example, record the type of interaction event (e.g., condom use) or the biological characteristics of the individuals that may affect the transmission probability. However, all simplifying assumptions in temporal networks are also used in the concurrency literature, so predictions based on concurrency cannot be more accurate than those made by temporal network epidemiology.

From a theoretical point of view, the largest conceptual simplification in the traditional concurrency literature is probably to use ``partnership'' as the unit transmitting the disease. Partnerships, as mentioned, are assumed to be capable of contagion throughout their duration. There is typically nothing in either questionnaires or theoretical papers that explicitly prevent a couple from resuming a partnership after they quit it~\cite{blanchard1990modelling,altmann1995susceptible,Morris1995SocNetw}. Without allowing partnerships to resume, the assumption that the transmission rate between a couple is time-invariant is probably a very coarse simplification of the data described by a temporal network. On the other hand, allowing partnerships to resume means that serial monogamy is no longer forcing time-respecting paths to be unidirectional (cf.\ Fig.~\ref{fig:Miller}). Furthermore, isolated sexual encounters (i.e., those occurring between two individuals just once) are typically excluded from questionnaires and mathematical/computational modeling for a few reasons. The first reason is that it may be too hard to remember the accurate timing of these encounters~\cite{Mah2010AidsBehav}. However, the reason why omitting an isolated sexual encounter would give more accurate results than to include it at the wrong time is unclear to us. The second reason is that these events would be negligible from the viewpoint of disease dynamics~\cite{Epstein2011JIntAidsSoc}. Reference~\cite{Epstein2011JIntAidsSoc} states that ``When the duration of concurrency is short, the connectivity of the networks is more transient, and less conducive to rapid spread.'' Notwithstanding, other papers argue occasional encounters~\cite{sawers_isaac,Foxman2006SexuallyTransmittedDiseases} and commercial sex~\cite{Inungu} are essential for the epidemics.

Gathering information about peoples' partnerships might be more feasible than registering their sexual encounters individually for any meaningful period. That is the only argument we can envision for building a theory on a concept as nebulous as ``partnership''. Still, there are data sets gathering individual encounters~\cite{contact_surveys}. If the community had explained the patterns of such data sets, there might not have been a decade-long theoretical dispute.  It seems necessary to agree on definition of how to construct a partnership edge from a temporal network composed of time-stamped events~\cite{Holme2013PlosComputBiol}. Although it would not affect medical epidemiology, it is expected to pave the way to standardize the too imprecise language of theorists.

While we can, and should, move forward with temporal network epidemiology, we are not arguing that we should forget the traditional concurrency theory. On the contrary, both the data gathered and the theoretical work done are so valuable that our job is to link them with temporal network theory. Furthermore, many people in the field are now so accustomed to discussing sexual contact patterns in terms of partnerships that it would be nearly impossible to erase their mental pictures, and abstracting a problem at different levels can often reveal different insights.

Concurrency is not the only idea that lingers in the scientific discourse despite newer concepts being more precise and informative. The ``basic reproductive number'' $R_0$ is another example from theoretical epidemiology that is both notoriously hard to estimate~\cite{dietz1993estimation} and ill-suited to parametrize theoretical models~\cite{Keeling2000JTheorBiol,HolmeMasuda2015PlosOne}. Another analogous situation is the idea that ecological networks tend to be ``nested''~\cite{MARIANI20191}, for which recent studies have pointed out that it might be more fruitful to think of nestedness as the consequence of more fundamental network structures~\cite{staniczenko2013ghost}. We believe that other concepts will eventually supersede concurrency, $R_0$, and nestedness, but there is no need to hurry that development.

How should we use and deepen the concept of concurrency? Is concurrency useful for analyzing and predicting real epidemic processes? Based on the discussion above and in the previous sections, we propose the following issues as something we should have in mind and understand better to harness the concept of concurrency.

Measures like $\kappa_3$ and its variants are, by construction, static. Of course, if one projects a temporal network to a static network at a time $t$, for example by constructing the momentary network, these measures will be functions of $t$ (cf.\ Fig.~\ref{fig:kappas}). Nonetheless, by being a static measure, we can analyze it by static network theory~\cite{Barabasi2016book,van2017random,Newman2018book}. This field has developed tremendously since the publication of the still most authoritative theoretical concurrency papers. Now we know that disease spreading depends not only on the node's degree~\cite{gupta1989networks,Anderson1991book,Abate1995OrsaJComput,Andersson2000book,Diekmann2000book,Pastorsatorras2001PhysRevLett,dezsHo2002halting,liljeros2003sexual,Barrat2008book} and the concurrency but also on, e.g., mesoscopic structures of networks~\cite{stegehuis2016epidemic,WU2008623} and the densities of short cycles~\cite{volz_miller_galvani_meyers,badham2010impact,ball2013network}. To properly evaluate concurrency by measures like $\kappa_{1}$ and $\kappa_{3}$, we need to integrate them in the broader theory of spreading phenomena on networks.

Furthermore, it is not only the case that other network structures than concurrency affect epidemics, but they also affect concurrency itself. In simple pair-formation models, such as the ones in Refs.~\cite{blanchard1990modelling,altmann1995susceptible,Morris1995SocNetw,KRETZSCHMARpairformation}, higher activity (i.e., node's degree) means more concurrency. Even though concurrency is a readily understandable concept and hence appropriate to communicate to the general public, the average number of events is even better. Although theoretical epidemiology was also the first field to study other network structures such as degree assortativity~\cite{gupta1989networks} or heterogeneous degree distributions~\cite{Anderson1986ImaJMathApplMedBiol,mollison1994epidemics}, strangely, the relationship between these structures is still not fully charted. Our understanding of how temporal structures are related to concurrency is even more limited~\cite{Moody2016AnnEpid,MillerSlim2017PlosOne,LeeMoodyMucha2019Springer,Holme2013PlosComputBiol,Holme2014SciRep,HolmeControlling2016,Leng2018Epid}. One such example is the statistics of the interevent times \cite{Masuda2013F1000,JoPerotti2014PhysRevX,Masuda2020PhysRevResearch,Karsai2018Springer,Holme2014SciRep}. One way of understanding temporal effects on concurrency measures, so far missing in the literature, would be to use randomized data sets as null models~\cite{Holme2005PhysRevE,Gauvin2020arxiv}.

A different feature of networks that impacts concurrency is higher-order interactions. Hypergraphs and simplicial complexes, in particular, enable us to represent interaction among more than two nodes, such as group conversations, in unified mathematical frameworks \cite{Lambiotte2019NatPhys,Battiston2020PhysRep}. Simultaneous interaction involving at least three nodes implies concurrency. Quantifying concurrency for hypergraphs and simplicial complexes as well as to use epidemic process models on these structures to study concurrency may be promising research directions.

Empirical studies of concurrency have almost exclusively focused on sexually transmitted diseases~\cite{Foxman2006SexuallyTransmittedDiseases,Aral2010CurrInfectDisRep,Lurie2010AidsBehav-limited,Kretzschmar2012AidsBehav,Sawers2013JIntAidsSoc}. Since sexual acts are clearly defined events, these diseases are appropriate to network epidemiology in general. Other infections such as influenza \cite{meyers2005network,salathe2010high} or COVID-19~\cite{Barrat2021JRSocInterface} also spread over networks, and concurrency should affect these as much as sexually transmitted ones. For these diseases, it is common to use other types of networks than those of individuals to represent contact structures. A node can represent a location (like a city, hospital ward), group of people, and so forth, which is an example of higher-order representations. Also in these models it may be fruitful to consider concurrency~\cite{liljeros2007contact,stopczynski2015physical,donker}. Assuming that a pathogen can linger at a location, it would be worse if two persons, A and B, visited it repeatedly and alternately (i.e., the location having high concurrency), than if A's last visit to the location preceded B's first visit.

In the last decade, our understanding of the structure of the networks on which infectious diseases spread improved tremendously~\cite{HolmeSaramaki2012PhysRep,Holme2015EurPhysJB,Karsai2018Springer}. Unlike sexual networks, it is the mobility of people~\cite{balcan2009multiscale,Schlosser202012326} that drives proximity networks. The term ``partnership'' is even more misleading for such mobility-induced networks, but the mechanism of concurrency as an accelerator of epidemics is probably still valid.

Whether concurrency is a useful target for mitigating disease spread is unclear, and may depend on the phase of the epidemic~\cite{MillerSlim2017PlosOne}. Decreasing concurrency without changing the total number of events would require that one proactively changes events' timing to reduce concurrent events that nodes experience. Such concurrency-based interventions are underexplored.

To summarize, there is much work left to unify the theory of concurrency with temporal network science. We need to agree on operational definitions of concepts like ``partnership'' and measures for both individual and system-wide concurrency. We also need to carry out more extensive studies to clarify whether concurrency is a sizable contributor to epidemic dynamics on networks compared to other static and temporal network properties.

\section*{Acknowledgments}
N.M. acknowledges support from AFOSR European Office (under Grant No.\ FA9550--19--1--7024), the Nakatani Foundation, and the Sumitomo Foundation. J.C.M. acknowledges the support of La Trobe University. P.H. was supported by JSPS KAKENHI Grant Number JP 21H04595. 


\end{document}